# Public archives for biological image data

## Authors

Jan Ellenberg[1*], Jason R Swedlow[2*], Mary Barlow[3], Charles E Cook[3], Ardan Patwardhan[3], Alvis Brazma[3*] & Ewan Birney[3*].

## Affiliations
1. European Molecular Biology Laboratory, Meyerhofstrasse 1, 69117 Heidelberg, Germany
2. Division of Computational Biology, School of Life Sciences, University of Dundee, Dundee DD1 5EH, UK
3. European Molecular Biology Laboratory, European Bioinformatics Institute (EMBL-EBI), Wellcome Genome Campus, Cambridge CB10 1SD, UK

* Authors for correspondence: jan.ellenberg@embl.de, jrswedlow@dundee.ac.uk, brazma@ebi.ac.uk, birney@ebi.ac.uk.

## Abstract
Public data archives are the backbone of modern biological and biomedical research. While archives for biological molecules and structures are well-established, resources for imaging data do not yet cover the full range of spatial and temporal scales or application domains used by the scientific community. In the last few years, the technical barriers to building such resources have been solved and the first examples of scientific outputs from public image data resources, often through linkage to existing molecular resources, have been published. Using the successes of existing biomolecular resources as a guide, we present the rationale and principles for the construction of image data archives and databases that will be the foundation of the next revolution in biological and biomedical informatics and discovery.

## Introduction

Since the mid-1970s it has been possible to analyze the molecular composition of living organisms: from the individual units (nucleotides, amino acids, metabolites) through to the macromolecules they are part of (DNA, RNA and proteins) and how these in turn can interact with each other.[1,2,3] The cost of such measurements has decreased remarkably, while technology development has widened their scope from measuring all gene and transcript sequences (genomics/transcriptomics) to proteins (proteomics) and metabolic products (metabolomics). However, we will not understand life just from knowing which molecules are present, we also need to know *when* and *where* they are present and how they interact inside cells and organisms. This requires mapping their spatiotemporal distribution, structural changes and interactions in biological systems.

Observation and measurement of the *when* and the *where* in living organisms pre-dates chemical measurements by over three centuries: such observation began with light microscopy[4] and was augmented centuries later with diverse "new" technologies, such as electron microscopy, X-ray imaging, electron beam scattering, and magnetic resonance imaging. These direct physical measurements range from atomic scale through to the whole organism and have a striking dual use: quantitative measurements of molecular structure, composition and dynamics across many spatial and temporal scales, and in parallel, images that provide powerful visual representations of biological structures and processes for the scientific community and the wider public. With the huge expansion of imaging at all levels enabled by revolutionary technologies including electron cryo-microscopy, volume electron, super-resolution light and light sheet microscopy, the opportunities for research and biomedical insight have never been greater. Delivering on this potential requires open sharing of image data to encourage both re-use and knowledge application as well as engagement with the public.

Despite the long history of biological imaging, the tools and resources for collecting, managing, and sharing image data are currently immature compared to those available for sequence and 3D structure data. "Imaging" is not a single technology, but an umbrella term for diverse technologies that create spatiotemporal maps of biological systems at different scales and resolutions (Table 1). The challenge is further complicated by the large size and diversity of image data sets and associated computational analysis tools. Achieving the level of integration for imaging data that is routine in biomolecular data will require the development of an information technology platform that harmonizes measurements across spatiotemporal scales, different imaging modalities, and data formats. This is a considerable challenge that, until now, has appeared too formidable and expensive to pursue.

Two developments now make it feasible and of utmost importance to address this challenge: breakthroughs in imaging technology and in computer science. First, the resolution revolution in light and electron microscopy (recognized with the Nobel prizes in chemistry in 2014[5], 2017[6]) now routinely allow molecular identification and structure determination in imaging data and rapid advances in automation and throughput allow to do so in a systematic manner. Similarly disruptive technical improvements are already foreseeable also for larger scale imaging technologies. Second, the emergence of new powerful computer-based image interpretation, quantification, and modelling algorithms have revolutionized our ability to process large image data sets and are underpinned by the rapid evolution of data storage and cloud computing technologies, with a concomitant decrease in technology costs. The establishment of integrated public bioimage data resources has therefore become feasible and highly valuable. Publicly available image data will facilitate scientific reproducibility and rigour, researcher and instrumentation productivity and will catalyze the emerging discipline of "bioimage informatics", just as freely available DNA sequence data stimulated the emergence of modern sequence centric bioinformatics.

**Box 1.**

> One prominent public imaging resource is in the area of structural biology. Increased sensitivity and new methods in electron microscopy have created an entirely new approach to atomic scale measurement of biological molecules, single-particle electron cryo-microscopy, which bridges the gap between the atomic resolution of crystallography and traditional light microscopy to provide visualisation of cellular structures at near-atomic resolution.
>
> The structural biology community has enthusiastically embraced these technologies, implementing resources for electron cryo-microscopy imaging (EMDB, http://emdb-empiar.org/, established at EMBL-EBI in 2002, and EMPIAR, http://empiar.org/, established at EMBL-EBI in 2014) robustly demonstrating the value of image resources[7, 8]. For electron microscopy the return on investment is immediately obvious: a single cryo-EM facility is estimated to cost £4,000 per day to run: long term storage of experimental outputs is clearly cost-effective.

**Data archives and added-value databases**

There are two distinct types of biological databases. *Data archives* are long-lasting data stores, with the dual goals of (1) faithfully representing and efficiently storing experimental data and supporting metadata, thus preserving the scientific record, and (2) making these data easily searchable and accessible to the scientific community. An archive serves as the authoritative public resource for the data, it does not aim to synthesize datasets or make value judgments beyond assuring adherence to standards and quality. Archives enable the connection of different datasets through common standardized elements, such as genes, molecules and publications. A typical example of a biological data archive is the European Nucleotide Archive[9] (ENA), which stores nucleotide sequences. Data archives often have a single global scope, even if they are provided by a distributed organization.

The second type, referred to as *added-value databases* or *knowledgebases*, are synthetic: they enrich and combine different datasets through well-designed analysis, expert curation and, where possible, meta-analysis. They typically provide integrated information and biological knowledge for users. An example of integrative analysis in the imaging domain is determination of the structure of macromolecular machines, which combines individual X-ray structures of the complex subunits and then integrates these with a cryo-EM map, thus suggesting a molecular mechanism of how the machine functions[10].

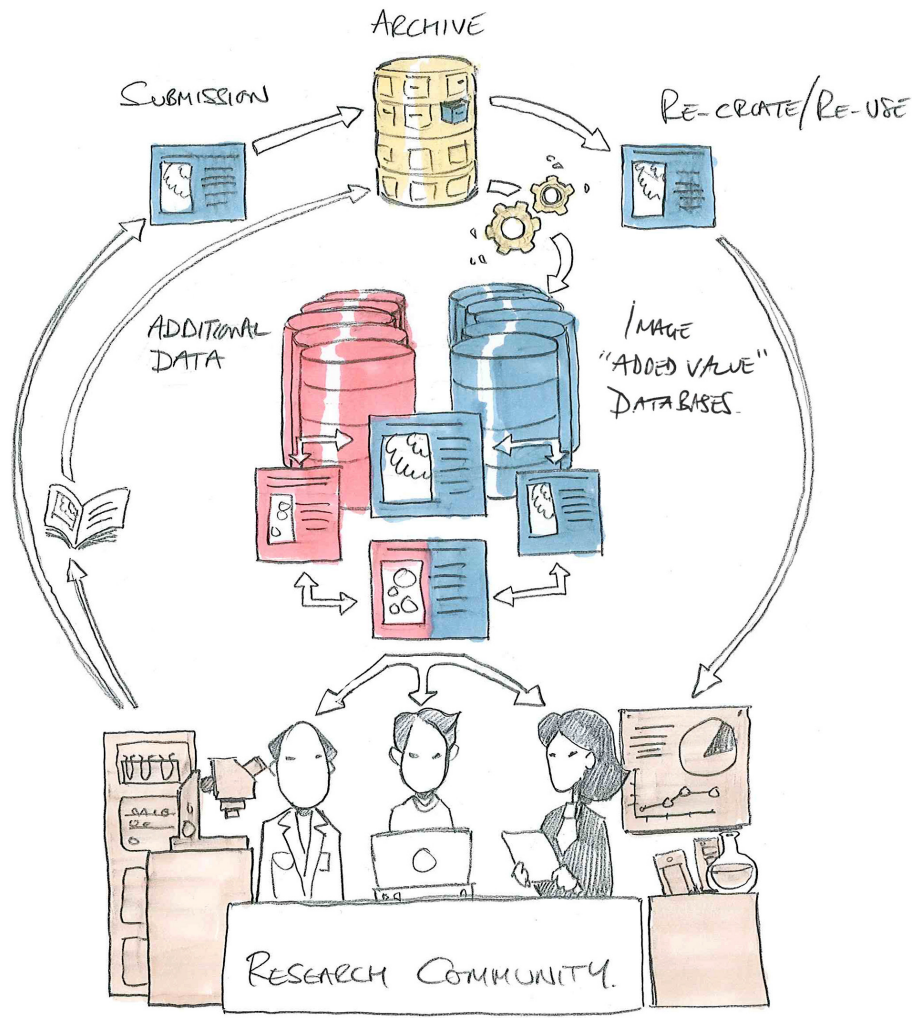

*Figure 1. The differing roles of data archives and added-value databases.*

A clear separation between data archives and added-value databases is important for ensuring the smooth flow of data from experimentalists generating the data into the public archives. For instance, if journals make it a requirement that data supporting a publication must be submitted to public archival databases, the data submission process should be straightforward, rapid, and should produce a unique identifier for citation of the data. Added-value databases may also want to apply careful curation to the submitted data, and may require additional quality control, updates to ontologies, or reprocessing of the data. The separation of the archival layer from the added-value layer makes both functions possible.

Thus, a functional biological imaging storage system would include both archival and added-value aspects. The archival repository of images would be preferably supported by an international collaborative effort with common standards in the same way that DNA archives are part of the International Nucleotide Sequence Database Collaboration consortium (INSDC, http://insdc.org)[9].

Added-value databases, developed around the archive, would focus on particular biological areas where greater understanding can be obtained by systematic integration of images, just as the Ensembl database and genome browser[11] add value to DNA data, and the Expression Atlas[12] integrates archived data to elucidate gene expression in a biology-centric way.

**The rationale for a bioimage archive**

The bioimage archive should store and make available imaging datasets from the molecular to organism scale (Table 1). Archived datasets should be directly related to the figures and other results included in a publication. Scaling of such an archive to the large number of publications that contain bioimage data is made possible by new, efficient, object-based storage systems, and would make data available for re-use, while also supporting integrity and reproducibility of research. A key aspect for sustainability of large-scale data archiving has been the deployment of technologies to ensure that the growth in volume of data year to year matches the decrease in disk storage costs; for DNA sequence this has required the development of data-specific compression.

To accelerate the development of the bioimage archive, and to demonstrate its value, datasets that are likely to be reused and are of high value to the community should be rapidly. *Reference images,* a concept that was introduced in a white paper published jointly by Euro-BioImaging and Elixir (http://www.eurobioimaging.eu/content-news/euro-bioimaging-elixir-image-data-strategy) are data that have value beyond a single experiment or project by also serving as a resource for the larger community. They must be interpretable by researchers outside the laboratory that generated them and also of general interest for many biologists. Examples of such images are systematic characterizations of proteins in a common cell type, mapping of spatio-temporal expression patterns or phenotypes or developmental atlases. At the beginning, defining the principles to identify datasets that are reference images will require a scientific review process. Over time however, image data sets are expected to attain a "highly used" label based on actual re-usage, similar to citation scores of scientific publications.

The scope of the archive would not be tied to any particular imaging technology, but rather defined by the likelihood of data reusability by the life sciences community. This is an established guiding concept in bio-molecular data archiving; for instance, the Gene Expression Omnibus (GEO) or the Functional Genomics Data Archive ArrayExpress[13, 14] accept data from different experimental modalities or generated by different technologies. The acceptance is based on the data's relevance to functional genomics, which in practice is often defined by the journal where the respective research paper is submitted, or the goals of the project that generated the data.

We list below four key synergistic functions that an integrated image data archive and associated added-value databases should fulfill:

a) *Open data and reproducibility* of analyses, allowing authors to provide a full audit trail of their original data and analysis methods, and allowing other interested scientists to explore alternative analyses of the same raw data. Making data, materials, and methods used in scientific research available to other researchers, with no usage restrictions other than requesting citation, is a long-standing tradition in the life sciences as well as an essential bedrock principle of scientific progress and scientific rigour[15, 16], http://opendatahandbook.org/guide/].
b) Provision of *reference data* for the research community. These data, provided with rules of community use and analysis methods, avoid the redundant production of replica datasets and can be the basis for added-value resources (e.g., atlases). They provide a large efficiency gain for routine experiments and over time will build a comprehensive scientific reference image resource for cells, tissues, organs, and organisms.
c) Allowing *new scientific discoveries* to be made using existing data, in particular via novel or integrated analysis of datasets that were originally generated for a different purpose or as a resource. The feasibility and value of such re-analysis of combined imaging datasets has already been demonstrated[17]. Integrative structure determination is another example of such a synergy.
d) Accelerating the development of *image analysis* methods against a broad range of benchmark datasets, thus encouraging validation and robustness of the computational methods and enabling well-structured image analysis challenges[18, 19, 20].

A similarly strategic approach to data archiving has been critical to the success of genomics—indeed it has amplified the value of DNA sequencing data enormously, as archives enable comparison of new sequences to all previously archived sequences. While we recognize that biological image data are more complex, context dependent, and multidimensional, such dependencies are not entirely new for molecular archives. Gene-expression data, for example, is also context dependent, multidimensional and subject to experimental conditions. Three-dimensional macromolecular structures and electron microscopy data have been archived for 45 and 15 years, respectively, and have proven to be tremendously valuable[21, 22, 23]. This previous experience suggests that the case for a central bioimage archive is extremely strong[7].

The experience of biomolecular data resources shows that the increased efficiency provided by reference data (function b) more than compensates for their storage and running costs, and that only open data enables assessment of reproducibility (function a). An integrated archive will provide a long-term home for image data that has high utility and reduces the need for every institution with significant image data-generation capabilities to invest in long-term archiving efforts. For consortia and projects planning to systematically generate large amounts of imaging data, the archive will become a trusted partner and help them to present cost-effective and sustainable solutions for image depositions and storage to their funders. As with biomolecular data resources, the maximum utility is realised when all scientists (whether academic or commercial) can freely deposit to the archive and freely access and re-use the data. Building and sustaining a public bioimage archive would bring substantial benefits across a wide range of scientific activities and domains.

There is a concern that a bioimage archive would be flooded by enthusiastic depositors offloading their storage costs to the central resource without scientific benefit beyond the submitting group. Our experience of running DNA archives and discussions with the community suggests this would not be the case. Nevertheless, to ensure control over costs and scale, if the storage capacity and/or staff effort of the archive are exhausted a light-weight independent expert assessment using the criteria defined above can be implemented.

**Easy submission of image data will drive the value of the archive**

One of the keys to successful sharing is the ease of submitting data to the archives. Submission criteria must balance collection of data and metadata necessary to interpret datasets against overburdening submitters with technical requirements. Moreover, it is not always obvious what metadata are minimally required, particularly when dealing with rapidly evolving technologies like imaging. The solution to this problem is for relevant scientific communities to rapidly agree on the initial minimum requirements and keep them up-to-date as technologies change and science advances. The underlying data structures will need to be flexible and to evolve with changing requirements, with submission tools co-evolving to keep up with these changes.

Additionally, journals and funding agencies should exert their influence to encourage data submissions. With evidence highlighting that enabling the sharing and re-use of data maximizes the impact of research, often increasing the volume of citations by one or two orders of magnitude[24,] the bioimage archive should work closely with journals and preprint servers to encourage image data deposition upon submission or acceptance for publication If the image data sharing infrastructure is to succeed it must demonstrate a positive return on the investment to both funders and data submitters. The involvement of funders will be crucial to ensure that large systematic data-generation projects commit to depositing their image data and appropriately budget for these costs within projects.

**Added-value image databases for knowledge extraction from the archive**

The bioimage archive will ensure the preservation of the scientific record while building up a critical mass of reusable data. The added-value databases will build on the archive to provide in-depth curation, annotation, standardization, reanalysis, and integration of independent datasets. They may also include advanced functionalities such as biological question-oriented searches and queries, cross-comparison of data sets, and advanced mining and visualisation. Examples of added-value image databases exist already, such as the Image Data Resource[17] (IDR, https://idr.openmicroscopy.org/) which integrates cell and tissue imaging studies based on genetic or drug perturbations and phenotype, and PhenoImageShare[25], which uses defined ontologies to integrate image datasets based on phenotype.

The bioimage archive should include pipelines to support existing and future added-value databases to maximize the possibilities of knowledge extraction and new discoveries. Given the currently available image data sets such as the EMDB, EMPIAR and IDR, and the rapid technical advances in image data generating technologies, we see exciting opportunities for expansion beyond already existing resources listed in Table 2 and an accelerated development of new resources.

**Human Image Archiving**

Many imaging modalities are used in clinical practice, and research and there are many synergies between clinical imaging and atomic-to-organism scale imaging (for example, robust data analysis is a common challenge). Due to the extensive use of imaging in clinical research and practice, there is also an increasing need for such imaging for human phenotyping and linked genotyping efforts; for example, in the UK BioBank[26]. Although there are analogies to human genome data sharing between researchers (e.g., as enabled by the European Genome-phenome archive, www.ebi.ac.uk/ega), data sharing in the clinical community has different challenges and a different culture. Additionally, practical factors such as appropriate consent, ethics approval, and data privacy must be addressed before broader sharing protocols can be set up. Our recommendation is to (a) continue to bring the atomic, cellular, tissue, and model organism imaging communities together with the clinical community and (b) continue a broad discussion on the merits and requirements of comprehensive data sharing among clinical researchers, with a particular focus on factors of known complexity, such as consent and privacy.

**Building a bioimage database system and its user communities**

In the future we envisage an integrated bioimage archive with easy submission systems and data standards interconnected with a growing and diverse set of added-value databases that together evolve into a comprehensive bioimage database system. The bioimage archive, like those already established for genomics, transcriptomics and proteomics, will be a large international effort, leveraging contributions, resources and technology from the global scientific community. In Europe, the ESFRI process has already provided a framework for the research infrastructure around imaging technologies, Euro-BioImaging (www.eurobioimaging.eu). During Euro-BioImaging's preparatory planning process representatives from almost all EU member states participated in user surveys, technological evaluations and proof-of-concept tests of a transnational bioimaging infrastructure. This activity produced plans for image data archiving and community-accessible tools for image analysis and processing. Euro-BioImaging has recently submitted an application to become a European Research Infrastructure Consortium (ERIC) to obtain independent international status and implement the plans developed in the preparatory phase.

Euro-BioImaging, its pilot image data resource IDR and the strategic collaboration with ELIXIR are examples of how national and transnational efforts can synergise and collaborate to deliver the components of the bioimage data ecosystem outlined here. Bioimage data resources will fuel life-science research not just in Europe, but globally. Indeed, connecting European efforts with international partners is a major objective of the Global BioImaging Project (www.eurobioimaging.eu/global-bioimaging).

**Conclusions**

Recent revolutionary advances in imaging technologies are providing increased quality, resolution and quantity of biological image data. With the rapid developments of computer-based image interpretation, quantification, and modelling algorithms, it is now feasible to process and derive value from large image data, opening up new opportunities for research, methods development and training, culminating in the acquisition of groundbreaking biomedical insights. To deliver on this promise the open sharing of biological image data is essential, allowing re-use and wider application.

We outlined above three developments that, together, create the opportunity to establish an open access Bioimage archive: 1) the development of new imaging technologies that generate large high quality image datasets; 2) faster computers and new computational methods the allow better and increased analysis of these images; and 3) the decreasing costs and increasing utility of data storage and cloud computing technologies. This archive will store reference images that will be freely available for re-use. "Reference" images will include any image that is formally published as well as other curated image datasets. The archive will, in turn, support the formation of a host of added-value image data resources that will enhance the scientific value of the archival images through curation and development of new analytical algorithms and methods.

The practical experience of the established resources shows that data volumes can be technically managed at scale and, importantly, can be governed in the future by clear and transparent criteria for data inclusion. We need to continue to expand engagement across the various bioimaging communities, in particular in clinical imaging. Nevertheless, we are already confident that broad bioimage archiving will deliver strong scientific and economic returns. The resulting basic research discoveries and clinical applications make the proposed integrated biological image resource (bioimage archive and the associated added-value databases) imperative for the biological and computational and in the future also clinical research communities.

**Acknowledgements**

We would like to thank the participants (R Ankerhold, Carl Zeiss Ltd; JA Aston, University of Cambridge; K Brakspear, MRC; JA Briggs, EMBL; LM Collinson, Francis Crick Institute; B Fischer, DKFZ; SE Fraser, University of Southern California; K Grünewald, Universität Hamburg; E Gustin, Janssen Pharmaceuticals; AN Kapanidis, University of Oxford; E Lundberg, KTH Royal Institute of Technology; RS McKibbin, BBSRC; TF Meehan, EMBL-EBI; SJ Newhouse, EMBL-EBI; AW Nicholls, GSK; DP Oregan, MRC; H Parkinson, EMBL-EBI; JW Raff, University of Oxford; Z Sachak, BBSRC; A Sali, University of California San Francisco; U Sarkans, EMBL-EBI; TR Schneider, EMBL; P Tomancak, MPI-CBG; J Vamathevan, EMBL-EBI) at an international workshop in BioImaging convened by the authors in Hinxton, UK, 23–24 Jan 2017 to explore the opportunities and challenges for biological image archiving. The goal of the workshop was to establish the scientific rationale for building public biological imaging data resources and the strategic principles that should guide their construction and much of the discussion above is distilled from dialogue at the workshop.

**References**

1. Maxam A.M. & Gilbert W. A new method for sequencing DNA. Proc. Natl. Acad. Sci. U. S. A. **74**, 560–564 (1977)
2. Sanger F & Nicklen A.R.C. DNA sequencing with chain-terminating. Proc. Natl. Acad. Sci. **74** 5463–5467 (1977)
3. Karas M, Bachmann D, Bahr U & Hillenkamp F. Matrix-Assisted Ultraviolet Laser Desorption of Non-Volatile Compounds. International Journal of Mass Spectrometry and Ion Processes **78** 53–68 (1987)
4. Leewenhoeck, A. Microscopical Observations, about Animals in the scurf of the Teeth, the substance call'd Worms in the Nose, the Cuticula consisting of Scales. Philosophical Transactions of the Royal Society **14**, 568–574 (1684)
5. "The Nobel Prize in Chemistry 2014". Nobelprize.org. Nobel Media AB 2014. Web. 8 Jan 2018. <http://www.nobelprize.org/nobel_prizes/chemistry/laureates/2014/>
6. "The 2017 Nobel Prize in Chemistry - Press Release". Nobelprize.org. Nobel Media AB 2014. Web. 8 Jan 2018. http://www.nobelprize.org/nobel_prizes/chemistry/laureates/2017/press.html
7. Patwardhan A. Trends in the Electron Microscopy Data Bank (EMDB). Acta Crystallographica.Section D, Structural Biology **73**, 503–508 (2017)
8. Iudin A., Korir P. K,, Salavert-Torres J,, Kleywegt G. J. & Patwardhan A. EMPIAR: A public archive for raw electron microscopy image data. Nature Methods **13**, 387–388 (2016)
9. Cochrane G., Karsch-Mizrachi I. & Takagi T. International Nucleotide Sequence Database Collaboration. The international nucleotide sequence database collaboration. Nucleic Acids Res **44**, D48–50 (2016)
10. Calloway E. The revolution will not be crystallized: A new method sweeps through structural biology. Nature **525**, 172-4 (2015)

11. Aken B. L., *et al.* Ensembl 2017. Nucleic Acids Res **45**, D635—D642 (2017)
12. Papatheodorou I, *et al.* Expression Atlas: gene and protein expression across multiple studies and organisms. Nucleic Acids Res 46, D246–251 (2018)
13. Clough E, & Barrett T. The Gene Expression Omnibus Database. Methods Mol Biol.**1418**, 93—110 (2016)
14. Kolesnikov N., *et al.* ArrayExpress update--simplifying data submissions. Nucleic Acids Res. **43**, D1113–6 (2015)
15. National Research Council (US) Committee on Responsibilities of Authorship in the Biological Sciences. Sharing publication-related data and materials: Responsibilities of authorship in the life sciences. Washington, D.C.: National Academies Press (2003)
16. Royal Society. Science as an open enterprise. London: The Royal Society Science Policy Centre (2012)
17. Williams E., *et al.* The Image Data Resource: A Bioimage Data Integration and Publication Platform. Nature Methods. **14**, 775—781 (2017)
18. Ludtke S. J., Lawson C. L., Kleywegt G.J., Berman H. M. & Chiu W. Workshop on the validation and modeling of electron cryo-microscopy structures of biological nanomachines. Pac Symp Biocomput **2011**, 369–373 (2011)
19. Ludtke S. J., Lawson C. L., Kleywegt G. J., Berman H. & Chiu W. The 2010 cryo-EM modeling challenge. Biopolymers. **97**, 651–654 (2011)
20. Marabini, R. *et al.* CTF challenge: result summary. J Struct Biol **190**, 348-359 (2015)
21. Berman, H. M., *et al.* The archiving and dissemination of biological structure data. Current Opinion in Sturctural Biology **40**, 17–22 (2016)
22. Lawson C. L., *et al.* EMDataBank unified data resource for 3DEM. Nucleic Acids Res.**44**, D396–403. (2015)
23. Patwardhan A., & Lawson, C. Databases and archiving for CryoEM. Meth Enzymol **579**, 393–412 (2016)
24. Wang, X., *et al.* The open access advantage considering citation, article usage and social media attention. Scientometrics **103**, 555 (2015)
25. Adebayo S., *et al.* PhenoImageShare: an image annotation and query infrastructure Journal of Biomedical Semantics **7**, 35 (2016)
26. Allen N. E., Sudlow C., Peakman, T., Collins R., & UK Biobank. UK biobank data: come and get it. Science Translational Medicine **6**, 224 (2014)
27. Sunkin S. M., *et al.* Allen brain atlas: An integrated spatio-temporal portal for exploring the central nervous system. Nucleic Acids Res **41,** D996–D1008 (2013)
28. Uhlén M., *et al.* Tissue-based map of the human proteome. Science **347**, 6220 (2016)
29. Thul P. J., *et al.* A subcellular map of the human proteome. Science **356**, 6340 (2017)
30. Cai Y, *et al.* An experimental and computational framework to build a dynamic protein atlas of human cell division. BioRxiv https://doi.org/10.1101/227751 (2017)
31. Ljosa V, Sokolnicki KL & Carpenter AE. Annotated high-throughput microscopy image sets for validation. Nature Methods **9**, 637 (2010)
32. Bray M-A, *et al.* A dataset of images and morphological profiles of 30 000 small-molecule treatments using the Cell Painting assay. Gigascience **22**, 1–5 (2017)
33. Neumann B, *et al.* Phenotypic profiling of the human genome by time-lapse microscopy reveals cell division genes. Nature. **464** 721-7 (2010)

34. Tohsato Y, Ho KHL, Kyoda K & Onami S. SSBD: a database of quantitative data of spatiotemporal dynamics of biological phenomena. Bioinformatics **32** 3471-3479 (2016)
35. Ring N, *et al.* A mouse informatics platform for phenotypic and translational discovery. Mammalian Genome **26** 413–421 (2015)

| Scale | Imaging technology | Use |
|---|---|---|
| **Molecular (Ångström)** | Single-particle cryo-EM (Electron Microscopy) and electron tomography averaging | Structural analysis, molecular function |
| **Molecular machines (nanometer)** | Cryo-EM, Super-resolution light microscopy (SRM) | Biochemistry, molecular mechanisms |
| **Cells (micrometer)** | Transmission EM, volume EM, light microscopy (widefield, confocal, SRM), electron tomography, 3D scanning EM, soft X-ray tomography | Cellular morphology, activity within cells, mechanism |
| **Tissues (millimeter)** | Volume EM, Scanning EM, light microscopy (multiphoton, light sheet, OPT, etc.), X-rays (microCT), fluorescence imaging, mass spectrometry imaging | Protein localisation, tissue morphology and anatomy, interactions between cells |
| **Organism/organ (centimeter)** | Photography, X-rays, magnetic resonance imaging (MRI), optical tomography technologies, computerized tomography (CT), luminescence imaging | Mechanistic understanding of development and disease |

**Table 1. Biological scales of imaging.**
Imaging is used to understand the behavior of organisms, the formation of embryos, the shape and dynamics of cells, and the structure, function and interactions of molecules that are the building blocks of life. There are many different imaging technologies and image types that work at different biological size and time scales, from the molecular to the whole ecosystem. In general, image capture at different scales uses different technologies and records different types of metadata.

| Data Type | Utility & Impact | Types of Users/Applications | Examples of Public Resources |
|---|---|---|---|
| **Correlative light and electron microscopy** | Link functional information across spatial and temporal scales | Structural biologists and Modellers: structural models that span spatial and temporal scales | EMPIAR[7] |
| **Cell and tissue atlases** | Construction, composition and orientation of biological systems in normal and pathological states | Educational resources; Reference for construction of tissues, organisms | Allen Brain atlas[27]; Human Protein Atlas[28]; Human Cell Atlas[29]; Mitotic Cell Atlas[30]; Model organism gene expression atlases[12] |
| **Benchmark datasets** | Standardised test datasets for new algorithm development | Algorithm developers; Testing systems | EMDataBank[7]; BBBC[31]; IDR[17]; CELL Image Library[32]; |
| **Systematic Phenotyping** | Comprehensive studies of cell structure, systems and response | Queries for genes or inhibitor effects; | MitoCheck[33]; SSBD[34]; IMPC[35]; PhenoImageShare[25] |

**Table 2. Examples of Potential High Value Datasets.**